\documentclass[%
 reprint,
showpacs,preprintnumbers,
 amsmath,amssymb,
 aps,
]{revtex4-1}
\usepackage{amssymb}
\usepackage{amsmath}
\usepackage{setspace}
\usepackage{url}

\usepackage{graphicx}
\usepackage{dcolumn}
\usepackage{bm}

\usepackage{subfig}
\usepackage[percent]{overpic}

\begin{document}


\title{Accelerating Growth and Size-dependent Distribution of Human Online Activities}

\author{Lingfei Wu}
\affiliation{%
 Department of Media and Communication, City University of Hong Kong.}%
\author{Jiang Zhang}
 \email{zhangjiang@bnu.edu.cn}
\affiliation{Department of Systems Science, School of Management, Beijing Normal University}%
\date{\today}

\begin{abstract}
Research on human online activities usually assumes that total activity $T$ increases linearly with active population $P$,
 that is, $T\propto P^{\gamma}(\gamma=1)$. However, we find examples of systems where total activity grows faster than active population.
 Our study shows that the power law relationship $T\propto P^{\gamma}(\gamma>1)$ is in fact ubiquitous
 in online activities such as micro-blogging, news voting and photo tagging. We call the pattern ``accelerating growth'' and
 find it relates to a type of distribution that changes with system size.
We show both analytically and empirically how the growth rate $\gamma$ associates with a scaling parameter $b$ in the size-dependent distribution.
As most previous studies explain accelerating growth by power law distribution, the
model of size-dependent distribution is novel and worth further exploration.

\end{abstract}


\maketitle


\section{Introduction}

There are two ways of describing the growth
of human online activities: linear and nonlinear. Linear
models assume that the average level of individual activities is a
constant. Hence, the total amount of
activity $T$ is a linear function of $P$, namely, $T\propto
P^{\gamma}(\gamma=1)$. For instance, the average degree is a
constant parameter in the Barab\'asi - Albert (BA) model \cite{s01}. However,
in nonlinear models the average level of individual activities increase
with system size \cite{s02}, leading to a power law relationship $T\propto
P^{\gamma}(\gamma>1)$. This relationship is supported by empirical studies on different online
activities including game playing \cite{s03}, resource
recommendation \cite{s04}, collaborative programming \cite{s05} and
tagging \cite{s06,s07}, as well as off-line activities such as energy consumption \cite{s08,s09,s10,s14}
and wealth creation \cite{s11,s12}.

Although there is plenty of evidence for nonlinear
growth (referred to hereafter as ``accelerating growth"), how such growth arises is still an open question.
In the current study, we find a power law relationship $T\propto
P^{\gamma}(\gamma>1)$ in serval types of online activities ranging
from micro-blogging, news voting to photo tagging. While previous authors have explained it with a
 long-tail distribution \cite{s05,s07,s16,s17,s18}, we propose that it is the ``size-dependent distribution",
rather than a long-tail distribution, that gives rise to
accelerating growth. It is observed that the behavior of highly
active users confirms to Zipf's law, while the behavior of less
active users does not. Therefore Zipf's law, also known as the rank
curve of power law distribution, fails to capture a regularity in
human online activities. Instead of Zipf's law, we use a discrete
generalized beta distribution (DGBD) \cite{s22} to fit the curves.
By tuning two parameters $a$ (which determines the activities of
highly active users and corresponds to the exponent $\alpha$ in
Zipf's law) and $b$ (which determines the activities of the less
active users) in the DGBD, we are able to fit the empirical curves
with $R^2 >0.9$. Furthermore, it is observed that the rank curves of
individual activities change with population size, and such a
correlation can be controlled by adding a scaling factor $P^{b}$ in
the DGBD function. We call the modified DGBD function the
``size-dependent distribution" and derive the relationship $T\propto
P^{\gamma}(\gamma>1)$ from it analytically. We therefore find that
the accelerating growth rate $\gamma$ is not determined by the
exponent in Zipf's law (or a power law distribution), as claimed by
previous studies \cite{s05,s07,s19,s21}, but is in fact related with
the size dependent exponent $b$.

\begin{center}
\begin{table*}[ht]\footnotesize
\caption{Estimates of accelerating growth.}
\small{(``N of Days'' is the number of observations of the given
dataset and "CI" stands for confidence interval.)}

\hfill{}
\begin{tabular}{lcccccccc}
\hline
  Activity &  Dataset & $\gamma$ & $95\% CI$ & Adjusted $R^{2}$ & N of Days & URL\\  \hline
  Photo tagging  & Flickr & 1.48 & [1.43, 1.54] & 0.93 & 193  & flickr.com\\
  Micro-blogging  & Jiwai & 1.19 & [1.03, 1.48] & 0.98 & 21 & m.jiwai.de\\
  News voting & Digg & 1.18 & [1.06, 1.22] & 1.00 & 31 & digg.com$/$news\\
  Book tagging & Delicious & 1.17 & [1.15, 1.19] & 0.93 & 663 & delicious.com\\   \hline
\end{tabular}
\hfill{}
\label{tab.1}
\end{table*}
\end{center}

\begin{figure}
\includegraphics[scale=0.5]{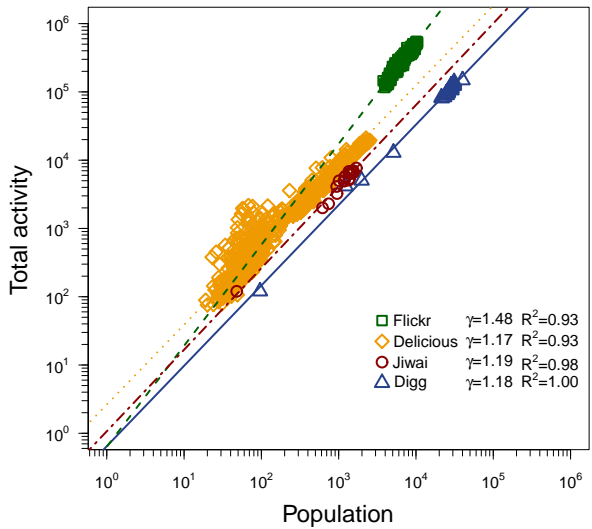}
\caption{(Color online) Accelerating growth in human online activities. Different
datasets are marked in points of different colors and shapes (green squares for Flickr, orange diamonds for Delicious, 
red circles for Jiwai, and blue triangles for Digg).
The x axis shows the active population in a day and the y axis shows the total activity in the day.
Both axes have a logarithmic scale. The orthogonal regression lines are also shown.}
\label{fig.1}
\end{figure}

\section{Accelerating growth in human online activities}

Our data cover several types of typical human online activities, including
the micro-blogging activities of 6,426 users on Jiwai, the news voting
activities of 139,409 users on Digg, the photo tagging activities of 195,575 users on Flickr,
and the book tagging activities of 13,988 users on Delicious. All the four data sets are publicly available.
The Jiwai data set is published in \cite{s23} and is available at
\url{http://www.fanpq.com/}. The Digg data set is published in \cite{s19} and can be downloaded from
\url{http://www.isi.edu/integration/people/lerman/downloads.html}. The Flickr and Delicious data sets
are published in \cite{s20} and can be downloaded from
\url{http://www.uni-koblenz-landau.de/koblenz/fb4/AGStaab/Research/DataSets/PINTSExperimentsDataSets/index_html}.
In the four systems, we define $P$ as the number of active users in a day, and $T$ the total
activity generated by these users. Note that the unit of $T$ is
different across systems: $T$ are micro-blogs in Jiwai, news votes in Digg,
and tags in Delicious and Flickr. The systems are summarized in Table~\ref{tab.1}.

We plot $T$ and $P$ in a log-log scale plot (Fig.\ref{fig.1})
and find a power law relationship
\begin{equation}
\label{eq.1}
{T}\propto{P}^{\gamma}.
\end{equation}
The values of $\gamma$ estimated by orthogonal regression are
shown in Table~\ref{tab.1}. We use orthogonal regression instead of ordinary
least squares regression because the latter tends to overstate
the effect of outliers \cite{s21}. The estimated values of $\gamma$ shown
in Table~\ref{tab.1} are all greater than 1. By noting that greater $\gamma$ means
more activities would be generated by a given population, we can regard $\gamma$ as
an indicator of productivity and compare the productivity between online and
off-line systems. Off-line activities, including wealth creation
and patent invention, have been found to scale linearly with population, $\gamma$ estimated to be
between 1 and 1.35 \cite{s09}. As one of the $\gamma$ in our study exceeds 1.40 (Flickr), this is evidence that
online systems could be more productive than off-line systems.

Our analysis of four online systems provides varying estimates of $\gamma$. But what determines the $\gamma$ ? In
exploring the underlying distribution of individual activities, we discover a new type of distribution that changes with
system size. In the following section, we introduce the size-dependent distribution,
which we propose relates to accelerating growth in human online activities. In particular, we show
how the exponent $b$ of the distribution can be used to predict the value of $\gamma$.

 \begin{center}
  \begin{figure*}[!ht]
   \centering
      \subfloat[\label{subfig-1:dummy}]{%
      \begin{overpic}[scale=0.5]{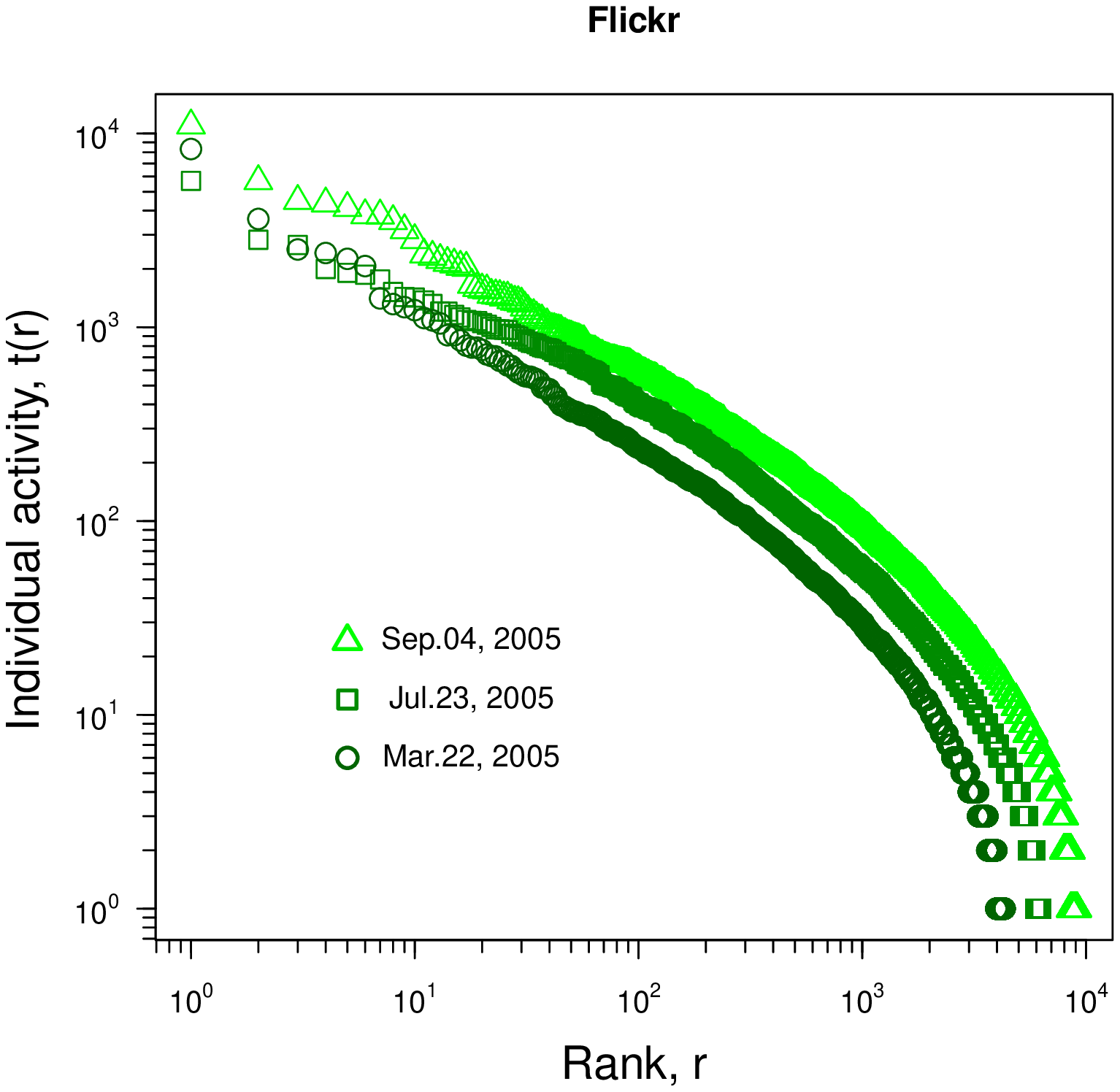}
        \put(60,55){{\includegraphics[scale=0.15]
{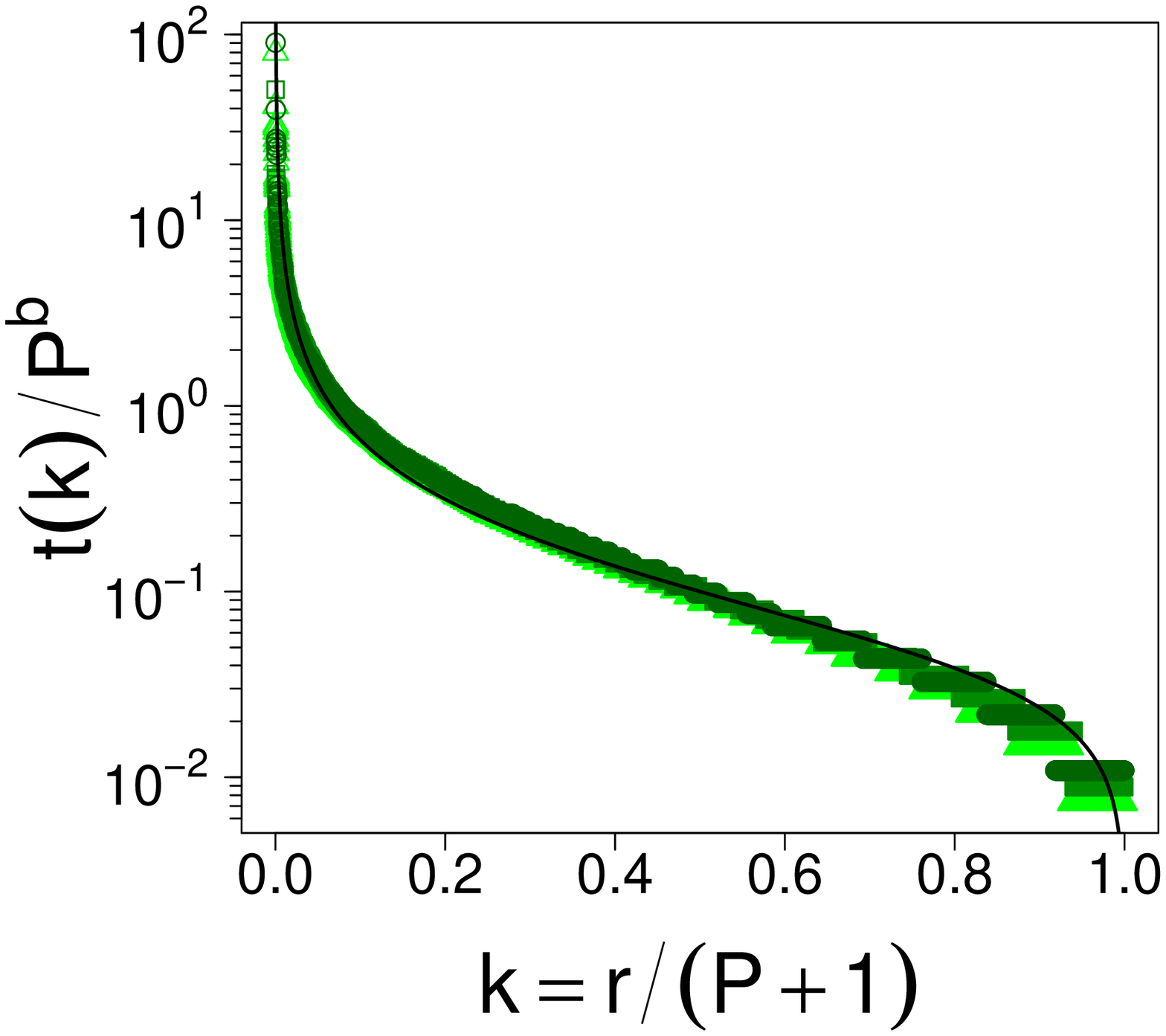}}}
      \end{overpic}
    }
       \subfloat[\label{subfig-2:dummy}]{%
      \begin{overpic}[scale=0.5]{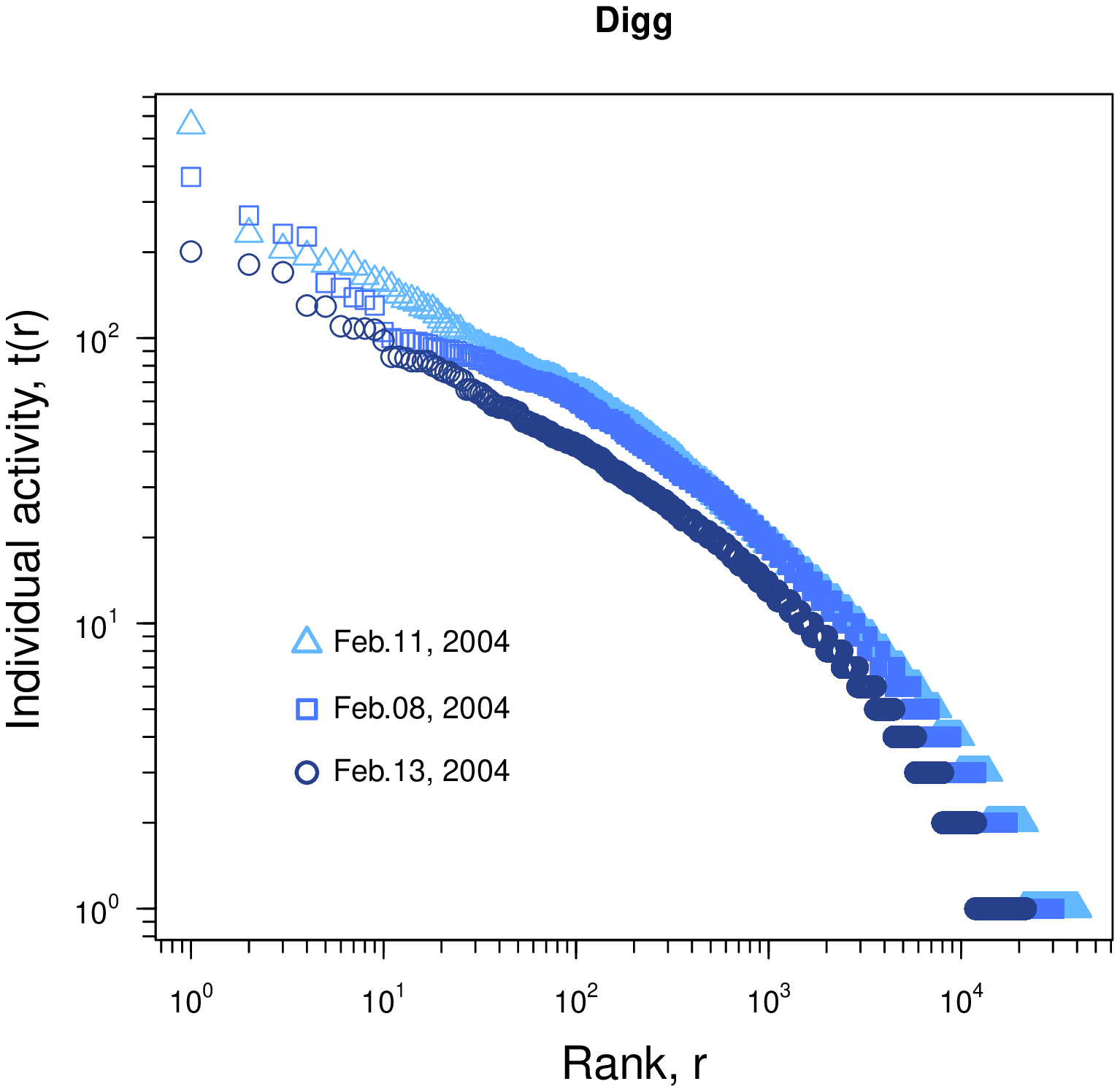}
        \put(60,55){{\includegraphics[scale=0.15]
{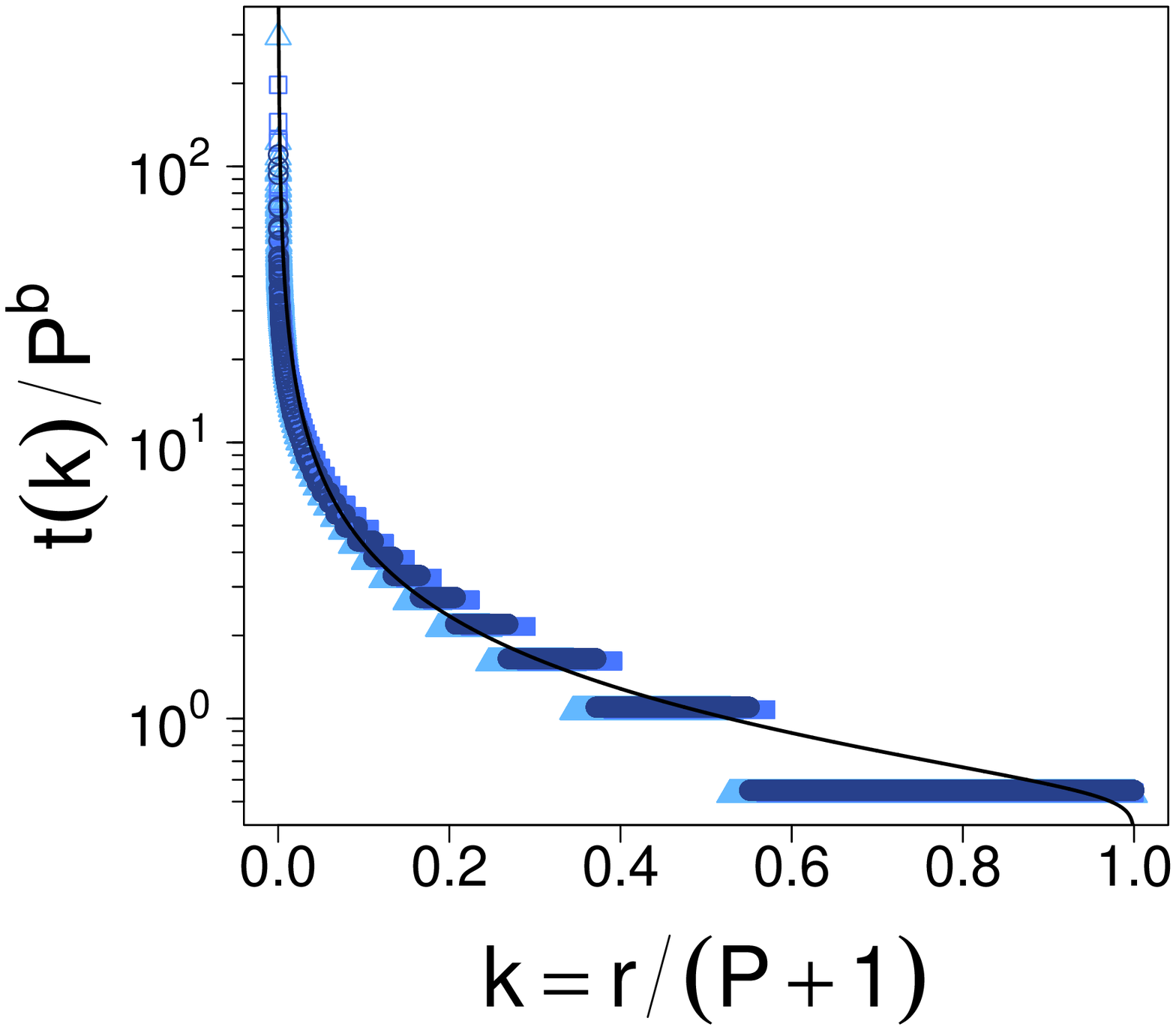}}}
      \end{overpic}
    }

    \centering
    \subfloat[\label{subfig-3:dummy}]{%
      \begin{overpic}[scale=0.5]{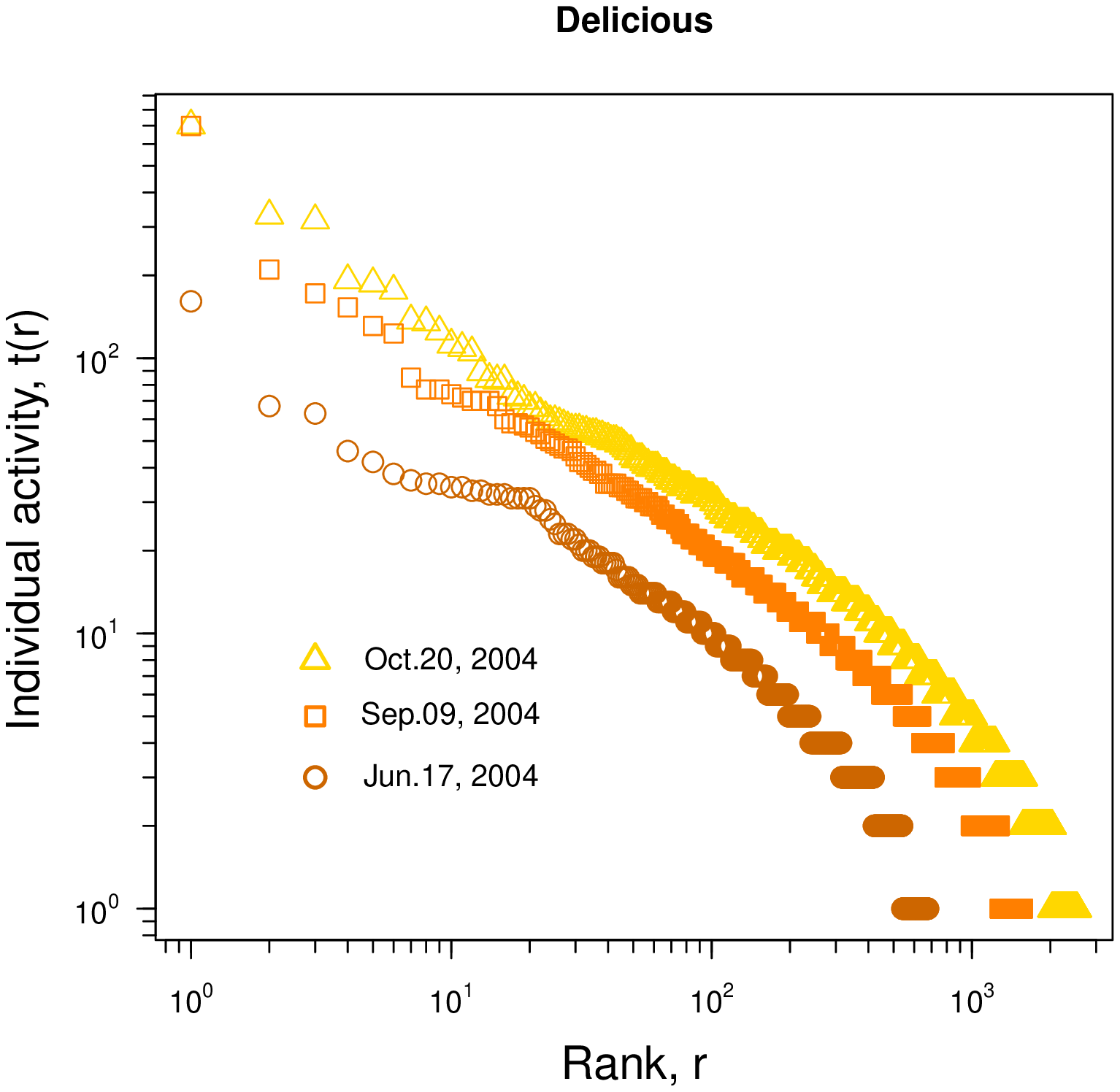}
        \put(60,55){{\includegraphics[scale=0.15]
{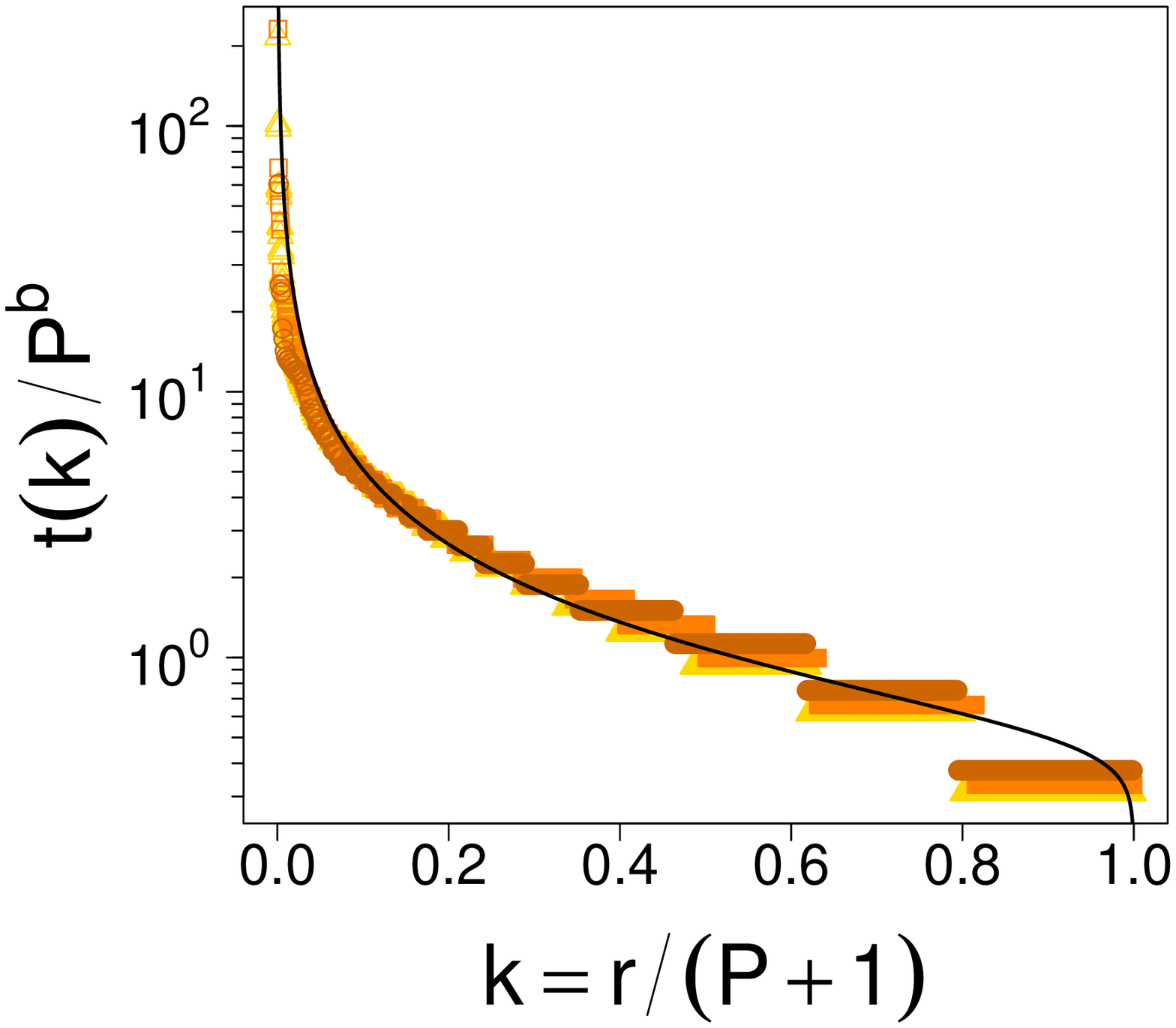}}}
      \end{overpic}
    }
    \subfloat[\label{subfig-3:dummy}]{%
      \begin{overpic}[scale=0.5]{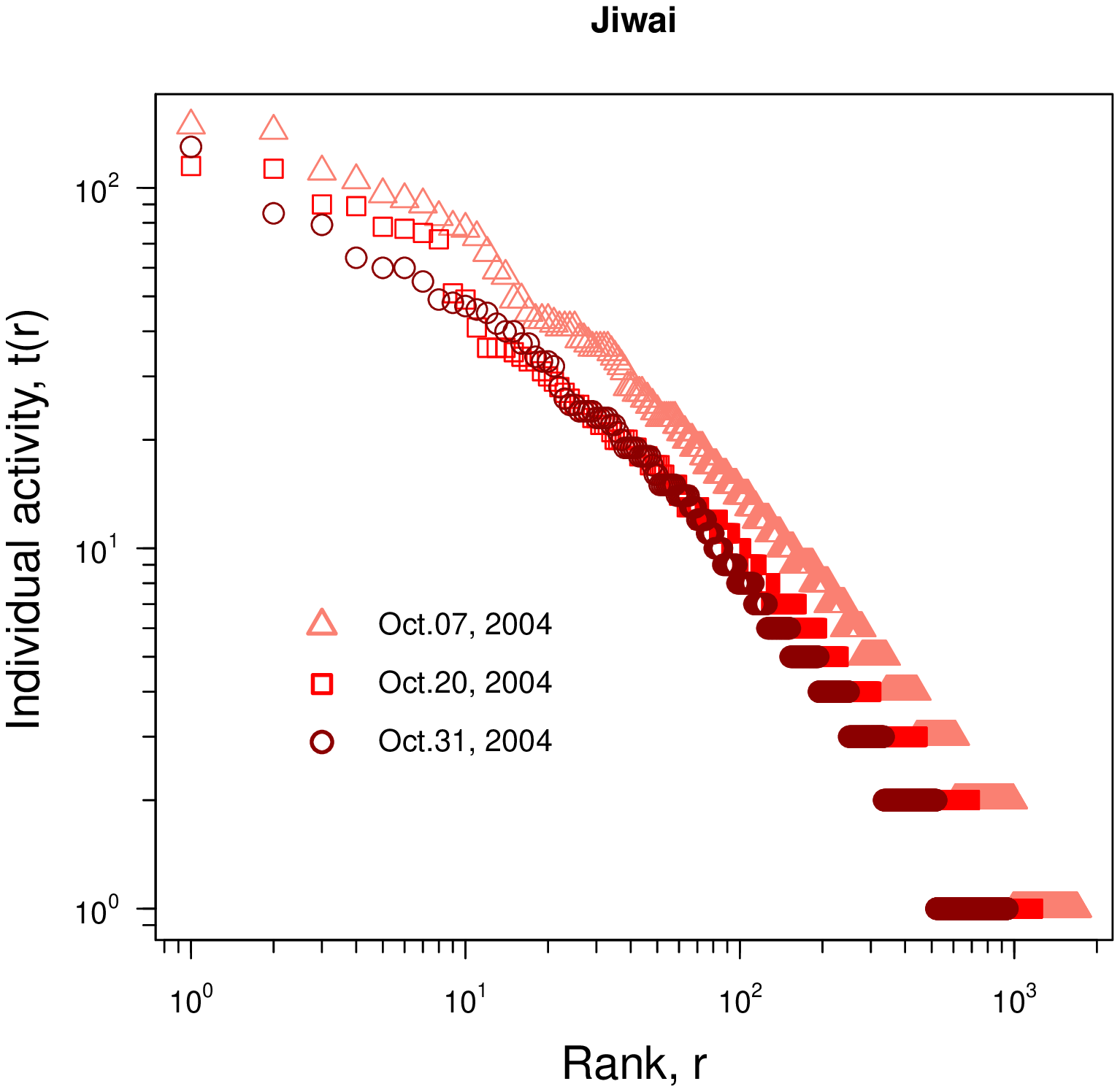}
        \put(60,55){{\includegraphics[scale=0.15]
{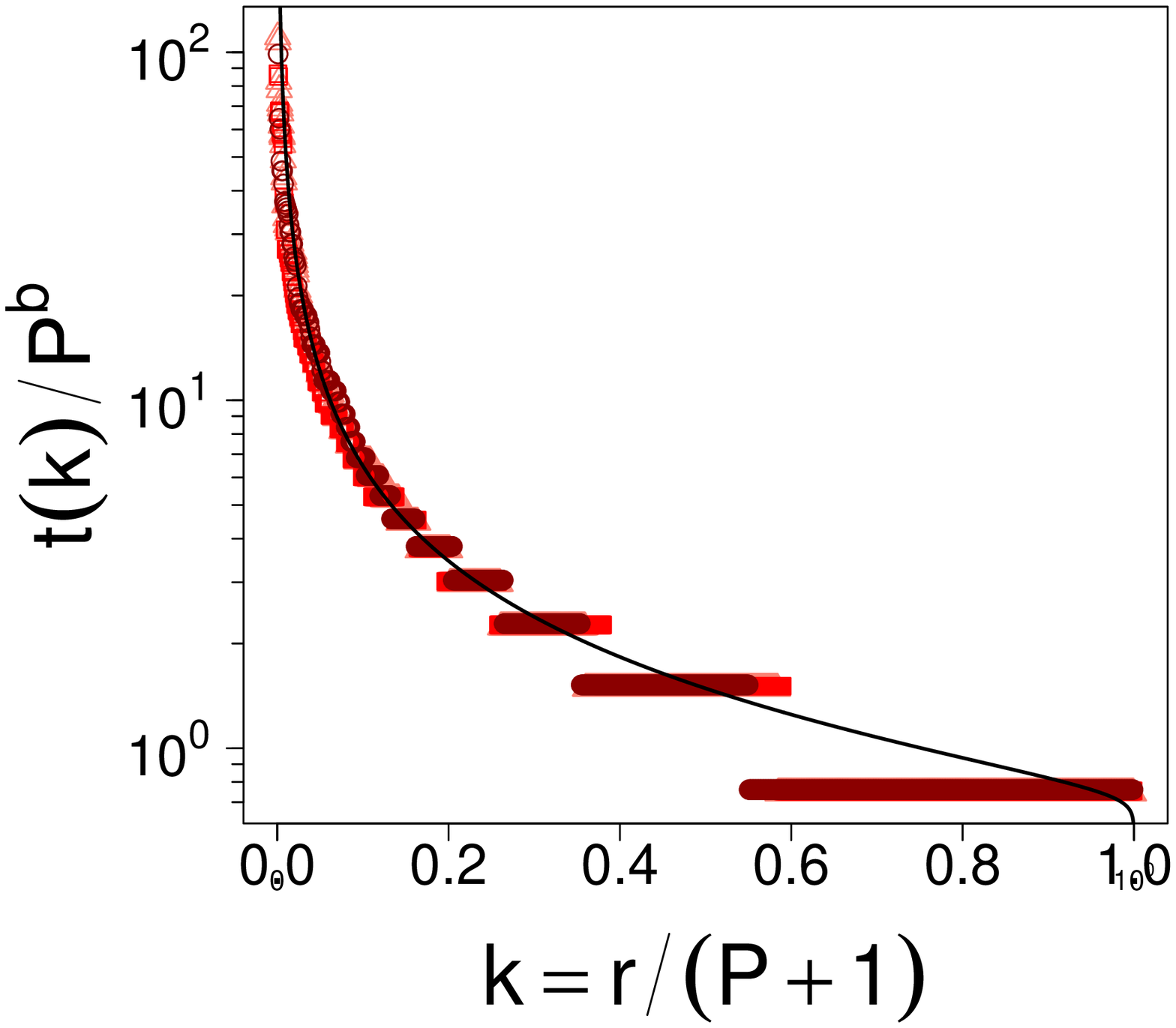}}}
      \end{overpic}
    }
    \caption{(Color online) Three examples of daily distribution of individual activities: Flickr (a),
    Digg (b), Delicious (c), and Jiwai (d).
     Different colors and shapes of the data points indicate different days. The y axis shows the individual activities and the x axis shows
     the decreasing ranks of the activities. Both axes have a logarithmic scale. The rescaled form of example distributions and the theoretical curves predicted by
      the size-dependent DGBD model are shown in semi-log plots (in which the y axis has a logarithmic scale) in the insets.}
       \label{fig.2}
    \end{figure*}
\end{center}

\section{\label{sec:level1}From size dependent distribution to accelerating growth}

In this section, we show that the DGBD, which has been extensively
studied for various data sets in \cite{s22}, fits the daily distributions of
individual activities in the four online systems. We then show that the distribution changes with system size and this
correlation can be captured by a scaling parameter $P^{b}$. In
other words, by adding $P^{b}$ to the distribution function, we can use it to
fit the empirical distributions in different days. We call the
modified distribution the size-dependent distribution and show that accelerating growth
in total activity can be derived from that distribution.

\subsection{\label{sec:citeref}The DGBD model}

We use $t(r)$ to denote the activity of a user in one
day, in which $r$ is the decreasing rank of the activity among all
individual activities in the day. Thus the maximum value of $r$, $r_{max}$, equals population
$P$. The DGBD model \cite{s22} of individual activities is then
\begin{equation}
\label{eq.2}{t(r)}=A(P+1-r)^{b}r^{-a}\quad (a>0, b>0),
\end{equation}
where $A$, $b$ and $a$ are parameters to be estimated. In
Fig.~\ref{fig.2}, we show three examples of daily distributions for
each system. Note that if we set $b=0$ then Eq.~(\ref{eq.2}) becomes
Zipf's law $t(r)\propto r^{-a}$. Therefore, the DGBD can be viewed
as a generalized Zipf's law (or power law distribution). The reason
for using the DGBD instead of Zipf's law, which has been widely used
to fit human behavioral data exhibiting a long tail
\cite{s12,s18,s22,s24}, is that in our data, the rank curves in the
rank-ordered plots (individual activity vs. rank) are not perfectly
straight lines. The empirical curves deviate from the straight line
predicted by Zipf's law at the right tails (Fig.\ref{fig.2}), and
hence estimations based on Zipf's law will be biased.

\subsection{\label{sec:citeref}The size-dependent DGBD model}

In fitting the daily distributions of individual activities with the DGBD,
we find that the parameter $A$ changes with population size. To confirm
this empirical finding, we analyze the DGBD function and find that
the relationship between $A$ and $P$ can be derived as follows.
As $r_{max}$ equals $P$ and the minimum value of individual activities $t(P)$
is 1 by construction (because we only consider active users),
we derive the boundary condition
\begin{equation}
\label{eq.3}t(r_{max})=t(P)=1.
\end{equation}
Substituting Eq.~\ref{eq.3} into Eq.~\ref{eq.2} gives
\begin{equation}
\label{eq.4} A = P^{a}.
\end{equation}
Let $k=r/(P+1)$. Obviously $k \in (0,1)$ and $1-k\in (0,1)$. By
replacing $r$ in Eq.~(\ref{eq.2}) with $k$ we get
\begin{equation}
\label{eq.5} t(k) = P^{a}(P+1)^{b-a}(1-k)^{b}k^{-a}.
\end{equation}
As in the data $P\gg 1$, namely, $P+1\approx{P}$, we can rewrite
Eq.~(\ref{eq.5}) as
\begin{equation}
\label{eq.6} t(k)\approx P^{b}(1-k)^{b}k^{-a},
\end{equation}
or
\begin{equation}
\label{eq.7} \ln(t(k))\approx b\ln(P(1-k))-a\ln(k).
\end{equation}
Eq.~(\ref{eq.6}) controls the variance of system size by replacing
$A$ with a scaling factor $P^{b}$, as well as normalizing rank $r$
into $k$. We refer to Eq.~(\ref{eq.6}) as the size-dependent DGBD
model and estimate its parameters $a$ and $b$ by ordinary least
squares regression (Table~\ref{tab.2}). The large adjusted $R^{2}$
is evidence that the size-dependent DGBD model captures the dynamic
properties of human online activities very well. Another way to
validate the size-dependent DGBD model is to plot $t(k)/P^{b}$ vs.
$k = r/(P+1)$ in different days and check whether the relationship
Eq.~(\ref{eq.6}) rises from data. The insets in Figure~\ref{fig.2}
shows that empirical distributions in different days collapse to the
same theoretical curves predicted by Eq.~(\ref{eq.6}), thus our
deduction is empirically supported.

\begin{table}\footnotesize
\begin{center}
\caption{Estimations of the size-dependent DGBD models.}
\begin{tabular}{lcccc}
\hline
  Dataset & $b$ & $a$ & Adjusted $R^{2}$ & N of Days \\ \hline
  Flickr & 0.54 &0.97 & 0.97 & 193 \\
  Jiwai  & 0.04 & 0.90 & 0.96 & 21 \\
  Digg  & 0.06 & 0.85 & 0.94 & 31 \\
  Delicious  & 0.15 & 0.91 & 0.94 & 663 \\ \hline
\end{tabular}
\label{tab.2}
\end{center}
\end{table}

We can derive the probability density function $f(x)$ of individual
activity from the size-dependent function Eq.~(\ref{eq.6}) as
follows. We know that the cumulative function $C(x)=Pr\{X>x\}$ of
the activity is the inverse function of the rank-activity curve
$t(k)$, namely,
\begin{equation}
\label{eq.8} C(x)=t^{-1}(x)=h(\frac{x}{P^b}),
\end{equation}
where function $h(x)$ is the inverse function of $(1-k)^bk^{-a}$,
that is:
\begin{equation}
\label{eq.9} h^{-1}(k)=(1-k)^bk^{-a}.
\end{equation}
The probability density function $f(x)$ can be derived from Eq.~(\ref{eq.8}) as
\begin{equation}
\label{eq.10}
f(x)=-\partial{C(x)}/\partial{x}=-\frac{1}{P^{b}}h'(\frac{x}{P^b}).
\end{equation}
Setting $g(x)=-h'(x)$ in Eq.~(\ref{eq.10}) gives a generalized form of the probability density
function:
\begin{equation}
\label{eq.11} f(x)=\frac{1}{P^{b}}g(\frac{x}{P^b}),
\end{equation}
which has already been found in off-line systems such as the stock market and the income distribution
\cite{baldovin_central_2007,stella_anomalous_2010,s12}.

\subsection{\label{sec:citeref}From the size-dependent DGBD model to accelerating growth}%
Accelerating growth can be derived from the size-dependent DGBD
model as follows. The integration of all user activities $t(r)$
is total activity $T$, that is,
\begin{eqnarray} \label{eq.12}
\nonumber T&=&\int_{r_{min}}^{r_{max}} \!t(r) \, \mathrm{d}r\\
 \nonumber&=&(P+1)\int_{0}^{1} \!t(k) \, \mathrm{d}k\\
 &\approx & P^{b+1}\int_0^1 \!(1-k)^bk^{-a} \, \mathrm{d}k.
 \end{eqnarray}
Using Euler integration, we can rewrite Eq.~(\ref{eq.12}) as
\begin{equation}
\label{eq.13} T\approx P^{b+1}\frac{\Gamma(1-a)\Gamma(1+b)}{\Gamma(2-a+b)},
\end{equation}
where ${\Gamma}$ is the gamma function. As $b$ and $a$ are
constants, according to the definition of the gamma function, we can
replace $\frac{\Gamma(1-a)\Gamma(1+b)}{\Gamma(2-a+b)}$ with a
constant $C$ and further rewrite Eq.~(\ref{eq.13}) as
\begin{equation}
\label{eq.14} T\approx CP^{b+1}.
\end{equation}
Eq.~(\ref{eq.14}) is the accelerating growth relationship. By comparing Eq.~(\ref{eq.1}) with Eq.~(\ref{eq.14}),
it is apparent that
\begin{equation}
\label{eq.15} \gamma\approx b+1.
\end{equation}
Note that the value of $\gamma$ only relates to $b$. As mentioned above,
Eq.~(\ref{eq.2}) becomes Zipf's law when $b=0$. Therefore
 Zipf's law leads to $\gamma=1$,
namely, linear increase of total activity with the growth of
population. Moreover, as it is the size-dependent parameter $P^{b}$
in Eq.~(\ref{eq.6}) that leads to accelerating growth, any distribution independent of
system size, including power law distribution predicted by the BA model \cite{s01}, can not result in accelerating growth.
As \cite{s22} reports the finding of the DGBD with a parameter $b>0$ in various empirical data sets,
it is reasonable to conjecture the wide existence of accelerating growth, as we have shown that $\gamma=b+1>1$.

To validate Eq.~(\ref{eq.15}), we can compare the theoretical and
empirical values of $\gamma$. Table~\ref{tab.3} shows $\gamma$
estimated from empirical data (from Table~\ref{tab.1}) and $\gamma'$
that is the theoretical value of $\gamma$ predicted by $b$ (from
Table~\ref{tab.2}). It is observed that the values of $\gamma$ and
$\gamma'$ are consistent with each other: all $\gamma'$ fall into
the $95\%$ CI of $\gamma$. Therefore our analytical deduction of the
relationship between size-dependent distribution and accelerating
growth is justified.

\begin{table}\footnotesize
\caption{The comparison between theoretical and empirical values of
$\gamma$.} \label{tab.3}
\begin{center}
\begin{tabular}{lcccc}
\hline Dataset & $\gamma$ within $95\% CI$ & $\gamma'$ & b & a \\
\hline
Flickr  & [1.43, 1.54] & 1.54 & 0.54 & 0.97 \\
Jiwai & [1.03, 1.48] & 1.04 & 0.04 & 0.90 \\
Digg  & [1.06, 1.22] & 1.06 & 0.06 & 0.85 \\
Delicious  & [1.15, 1.19] & 1.15 & 0.15 & 0.91 \\ \hline
\end{tabular}
\end{center}
\end{table}

It should be noted that the generalized form of the size-dependent
probability density function Eq.~(\ref{eq.11}) is a sufficient
condition of accelerating growth, meaning that if we replace $g(x)$
with other functions, we can still obtain the power law relationship
between system size $P$ and total activity $T$ \cite{s12}. This
finding is consistent with our previous study on income
distributions of countries\cite{s12}.

\section{\label{sec:level1}Conclusions}

In this paper, we discuss accelerating growth in human
online activities, that is, a power law relationship between total
activity and active population with an exponent greater than 1. The
power law relationship is found to be ubiquitous across different types
of human online behaviors. We show analytically how
size-dependent distribution relates to
accelerating growth, and validate our deduction using several large data sets containing
millions of human online activity records.

The major theoretical contribution of this paper is the finding that
size-dependent distribution relates to accelerating growth
quantitatively. Although our study is based on human online
activities, this quantitative relationship is not necessarily
confined to an online context. The model may also be used to explain
accelerating growth patterns in off-line social systems such as
cities \cite{s09} and countries \cite{s11,s12}.

Beside the theoretical contribution, the model of size-dependent distribution
has potential applications, e.g., in web crawling and website management.
For example, with historical data on individual activities,
webmasters can estimate the value of $b$ and predict the
accelerating growth rate $\gamma$ of total activity, which may help webmasters
plan the capacity of web server accordingly. Webmasters
can also compare the values of $\gamma$ among websites with equivalent functions,
leading to an innovative theoretically informed approach of benchmarking.

It should be noted that our findings appear to contradict conclusions of previous
studies. For example, \cite{s05,s07,s19,s21} suggest that the exponent $\gamma$ in accelerating growth
is determined by the exponent $\alpha$ in Zipf's law, but our analysis suggests that a power law distribution that is independent
of system size will not lead to accelerating growth. We conjecture
this contradiction may be due to an unknown relationship between the parameters $a$ (which, as mentioned,
corresponds to the exponent $\alpha$ in Zipf's law) and $b$ in the DGBD model,
since we have shown that $\gamma=b+1$. The unknown relationship between $a$ and $b$, together with other unsolved problems
such as the behavioral origins of size-dependent distributions, call for further exploration.

\begin{acknowledgements}
The authors thank Jonathan J. H. Zhu and Robert Ackland for providing comments on an earlier version of this paper.
One of us (J.Z.) acknowledges the support from the National Natural Science Foundation of China
under Grant No. 61004107.
\end{acknowledgements}

\bibliography{test}

\end{document}